\documentclass[preprint,12pt]{elsart}

\usepackage{amssymb,amsmath,cite}
\usepackage{graphicx,epstopdf,epsfig,color}
\def\bq{\begin{equation}}
\def\ee{\end{equation}}

\def\f{\frac}
\def\lt{\left}
\def\rt{\right}
\def\md{\,\mbox{d}}

\def\bk{k_{\mbox{\tiny B}}}
\def\sT{\sigma_{\mbox{\tiny T}}}

\newcommand\lae[1]{\label{#1}}

\journal{Physica A}

\begin{document}

\begin{frontmatter}

\title{Modelling of transport phenomena in gases based on quantum scattering}

\author{Felix Sharipov}
\ead{sharipov@fisica.ufpr.br}
\ead[url]{http://fisica.ufpr.br/sharipov}

\address{Departamento de F\'\i sica, Universidade Federal do Paran\'a, Curitiba, 81531-980 Brazil}

\begin{abstract}

A quantum interatomic scattering is implemented in the direct simulation Monte Carlo (DSMC) method applied to transport phenomena in rarefied gases. In contrast to the traditional DSMC method based on the classical scattering, the proposed implementation allows us to model flows of gases over the whole temperature range beginning from 1 K up any high temperature when no ionization happens. To illustrate the new numerical approach, two helium isotopes $^3$He and $^4$He were considered  in two canonical  problems, namely, heat transfer between two planar surfaces and planar Couette flow.
{To solve these problems, the} {\it ab initio} {potential for helium is used, but the proposed technique can be used with any intermolecular potential.}
The problems were solved over the temperature range from 1 K to 3000 K and for two values of the rarefaction parameter $\delta=1$ and 10. The former corresponds to  the transitional regime and the last describes the temperature jump and velocity slip regime.  No influence of the quantum effects was detected within the numerical error of 0.1 \% for the temperature 300 K and higher. However, the quantum approach requires less computational effort than the classical one in this temperature range. For temperatures lower than 300 K, the influence of the quantum effects exceed the numerical error and reaches 67\% at the temperature of 1 K. 
\end{abstract}


\end{frontmatter}

\section{Introduction}

The direct simulation Monte Carlo (DSMC) method  \cite{Bir12} used to calculate rarefied gas flows consists of decoupling of the free motion of gaseous molecules from collisions between them. The second stage requires a physical intermolecular potential in order to obtain reliable results. Recently, a procedure to implement any potential into the DSMC method was proposed in our previous paper \cite{Sha90} using the phenomenological Lennard-Jones potential as an example. In contrast to phenomenological models, {\it ab initio} (AI) potentials are free from any adjustable parameter usually extracted from experimental data. Nowadays, such potentials practically for all noble gases and their mixtures are available in the open literature, see e.g. \cite{Prz01,Cyb01,Bar18,Hel02,Hel03,Jag02,Jag03,Cac01}. Thus, the DSMC method  based on AI potential \cite{Sha96} also becomes free from such adjustable parameters. The idea of the procedure to implement any potential into the DSMC is to generate look up tables of the deflection angle depending on the relative velocity  of interacting particles and their impact parameter. The method was used to study the influence of the interatomic potential on various phenomena in rarefied gases \cite{Sha100,Sha105,Sha115,Sha116,Sha117} considering the intermolecular interaction  based on the classical mechanics, that is justified at high temperatures for heavy gases. However, the quantum effects in intermolecular interactions is not negligible for light gases, e.g. helium, hydrogen, tritium, especially at moderately low temperatures \cite{Joa01,Lan12,Hir01,Cha04,Fer02}.  It can be important to model helium, hydrogen and tritium flows in many technological fields such as cryogenic pumps \cite{Tan10,Fre04}, cryogenic systems used in the huge fusion reactor ITER \cite{Day02,Zha06}, monochromatic beams of helium \cite{Toe01,Ede02}, helium microscope \cite{Pal01,Bar19}, acoustic thermometry at a low temperature \cite{Pit05,Fis01}, experimental set-up to measure the neutrino mass \cite{Sha61,Sha72}, etc. In spite of the high practical interest to model gases at low temperatures, the quantum scattering has not been implemented yet in the DSMC method.

The aim of the present paper is to propose a new technique to implement the quantum scattering into the DSMC method {using any potential} and to show the influence of quantum effects on transport phenomena in rarefied gases. For this purpose, a procedure of generation of deflection angle matrix based on quantum scattering is elaborated and a couple of classical problems of fluid mechanics is solved to evaluate the influence of quantum effects. A temperature range where the classical approach fails and the quantum theory becomes an unique alternative to simulate the transport phenomena in rarefied gases will be pointed out. It will be also shown that even at a high temperature when the classical approach works, the quantum approach reduces computational effort that makes it preferable for the whole range of the temperature. It should be emphasized that we are interested in quantum effects only in interatomic iterations. Other effects, like high densities at low temperatures when the interatomic distance is comparable to the de Broglie wavelength, are not considered here.

\section{Numerical method}

The DSMC method consists of a decoupling the free-motion of molecules from  intermolecular collisions during each time steps $\Delta t$. Here, the free-motion of particles is considered to be classical that is valid under the condition \cite{Pat01,Man07}
\begin{equation}
\f{nh^3}{(2\pi m\bk T)^{3/2}}\ll 1,
\lae{DG}\end{equation}
where $n$ is the gas number density, $h$ is the Planck constant, $m$ is the atomic mass of the gas, $\bk$ is the Boltzmann constant and $T$ is the gas temperature. This condition is well satisfied at the atmospheric pressure and at any temperature above the boiling point of both $^3$He and $^4$He that are 3.2 K and 4.2 K, respectively. For a temperature below the boiling point, the gas pressure must be low enough to keep helium in the gaseous phase and to meet the condition (\ref{DG}).

In the present paper, the quantum scattering is implemented in  the stage of the intermolecular interactions. According to the no-time-counter version of the DSMC method \cite{Bir12}, the number of pairs to be tested for collisions during a time step $\Delta t$ in a cell of volume $V_c$ reads
\begin{equation}
N_{\mbox{\scriptsize coll}}=\f12 N_p(N_p-1)F_N (\sT g)_{\mbox{\scriptsize max}}\f{\Delta t}{V_c},
\lae{CL}\end{equation}
where $N_p$ is the number of particles in the cell, $F_N$ is the number
of real particles represented by one model particle, $g$ is the relative speed of two interacting particles, $\sT$ is the total cross section (TCS) of particles which is a function of $g$ according to the quantum theory of scattering,  the quantity   $(\sT g)_{\mbox{\scriptsize max}}$ represents a maximum value of the product $\sT g$ in each specific cell. Then, $N_{\mbox{\scriptsize coll}}$ pairs within the cell are chosen randomly. If a selected pair of particles satisfies the condition
\begin{equation}
\sT g/(\sT g)_{\mbox{\scriptsize max}}>R_f,
\lae{CQ}\end{equation}
the post-collision velocities are calculated; otherwise, the pre-collision velocities are kept. Here, $R_f$ is a random fraction varying uniformly from 0 to 1. The relation of the post-collision velocities to pre-collision ones contains the deflection angle
$\chi$ and impact angle $\varepsilon$, see Eqs.(8.32) - (8.35) from Ref. \cite{Sha02B}. The angle $\varepsilon$ is chosen randomly from the interval [0,2$\pi$], while the deflection angle $\chi$ should be calculated using the differential cross section (DCS) $\sigma(g,\chi)$ determined by the relative speed $g$. In contrast to the classical scattering used in the previous works \cite{Sha90,Sha96,Sha100,Sha105,Sha115,Sha116,Sha117}, here the DCS is needed not only to calculate the post-collision velocity, but even to test a pair before to accept or to reject it.

Exact calculations of the DCS in the frame of quantum theory is a hard task and it is completely infeasible to do such calculations for each tested pair. To avoid all this job, look-up tables of the deflection angle $\chi$ generated once for some specific gas can be used for any flows of this gas. The authors of Ref. \cite{Sos01} also proposed to store the incomplete cross section and then they simulated a spatially homogeneous relaxation of helium in several background gases. However, they used an equally spaced mesh of the deflection angle that requires a rather long procedure to generate this angle for each binary collision. To reduce this effort significantly in the present work, the matrix of the deflection angles is generated by such a way that all its elements for some specific speed $g$ are equally probable and can be chosen just randomly.

If we denote the DCS for the relative speed $g$ as $\sigma(g,\cos\chi)$, then the TCS is calculated as
\begin{equation}
\sT(g)=2\pi\int_0^\pi \sigma(g,\cos\chi)\sin\chi\md\chi.
\lae{CK}\end{equation}
The TCS also can be calculated directly without knowledge of the DCS. The method to calculate both DCS and TCS for some specific potential is given in Section \ref{SecA}. To implement it into the DSMC, we discretize  the speed $g$ introducing a sequence of nodes $g_j$ ($1\le j\le N_g$). They can be  distributed either equally spaced  or non-uniformly. It is important to have a simple rule to calculate a node $g_j$ which is nearest to the real speed $g$ of a pair chosen for collision. First, the TCS is calculated for each node $g_j$, i.e. $\sigma_{\mbox{\tiny T}j}=\sT(g_j)$. Then, the incomplete cross section is defined for each node of $g_j$ as
\begin{equation}
W_j(\cos\chi)=\f{4\pi}{\sigma_{\mbox{\tiny T}j}} \int_{\chi}^{\pi/2}  \sigma(g_j,\cos \chi')\sin\chi' \md\chi'.
\lae{AG}\end{equation}
In case of a single gas, it is enough to consider the angle range $0\le\chi\le\pi/2$ because the DCS based on the quantum scattering of undistinguishable particles is always symmetric $\sigma(g,\cos\chi)=\sigma(g,\cos(\pi-\chi))$.
{In case of mixtures, interacting particles are distinguishable so that the whole range $0\le\chi\le\pi$ should be taken into account.}
It is more convenient to express the incomplete cross section in terms of the new variable $\xi=\cos\chi$ as
\begin{equation}
W_j(\xi)=\f{4\pi}{\sigma_{\mbox{\tiny T}j}}
\int_{0}^\xi  \sigma(g_j,\xi') \md \xi',\quad 0\le\xi\le 1,
\lae{AH}\end{equation}
so that the matrix of scattering angles $\chi_{ij}$ can be represented by the matrix of their cosines $\xi_{ij}$. Note that $0\le W_j(\xi)\le 1$ as a consequence of (\ref{CK}) and (\ref{AG}).  To generates the matrix $\xi_{ij}$  with $N_\xi$ equally probable elements for each $j$th row, the following recurrent rule is used
\begin{equation}
W_j(\xi_{i+1,j})=W_j(\xi_{ij})+\f{1}{N_\xi},\quad 1\le i\le N_\xi-1,
\lae{CI}\end{equation}
where the first node for each speed $g_j$ is calculated as  $W_j(\xi_{1j})={1}/(2N_\xi)$.

The main "defect" of the classical approach is that the calculated TCS becomes infinite because theoretically two classical particles interact with each other at any distance $d$ between them. To avoid such a non-physical behaviour, the intermolecular potential must be cut-off, i.e. we assume that two particles do not interact with each other when the distance $d$ between them exceed some limit quantity $d_m$. In this case, the TCS is constant and equal to $\sT=\pi d_m^2$. Then the expression (\ref{CL}) becomes
\begin{equation}
N_{\mbox{\tiny coll}}=\f12 N_p(N_p-1)F_N \pi d_m^2 g_{\mbox{\scriptsize max}}\f{\Delta t}{V_c},
\lae{CS}\end{equation}
while the condition (\ref{CQ}) is reduced to $g/g_{\mbox{\scriptsize max}}>R_f$. The quantity $d_m$ should be sufficiently large so that its further increase could not change results of simulation within an adopted error. The technique to calculate the deflection angle matrix in the frame of the classical theory is quite different and can be found in Ref. \cite{Sha90}.

\section{Differential cross section}\lae{SecA}

According to the quantum theory of scattering \cite{Joa01,Lan12,Hir01,Cha04,Fer02}, the  DCS of undistinguishable particles with a spin $s$ consists of two terms and reads
\begin{equation}
\sigma^{\mbox{\tiny (B)}}(g,\chi)=
\f{s}{2s+1}\sigma'(g,\chi)+
\f{s+1}{2s+1}\sigma'' (g,\chi),
\lae{CH}\end{equation}
\begin{equation}
\sigma^{\mbox{\tiny (F)}}(g,\chi)=
\f{s+1}{2s+1}\sigma'(g,\chi)+\f{s}{2s+1}\sigma'' (g,\chi),
\lae{CZ}\end{equation}
for boson and fermions, respectively. Both $\sigma' $ and $\sigma''$ are expressed via the speed $g$ and deflection angle $\chi$ as
\begin{equation}
\sigma'(g,\chi)=\f{2}{k^2}\lt|\sum_{l=1,3,5,...}^\infty f_l(g,\chi) \rt|^2,
\lae{CT}\end{equation}
\begin{equation}
\sigma'' (g,\chi)=\f{2}{k^2}\lt|\sum_{l=0,2,4,...}^\infty f_l(g,\chi) \rt|^2,
\lae{BR}\end{equation}
\begin{equation}
f_l(g,\chi)=(2l+1)
\exp\lt(i\delta_l\rt)\sin \delta_l  P_l(\cos\chi),
\lae{DA}\end{equation}
with the only difference is that $\sigma'$ has only odd $l$, while  $\sigma'' $ contains only even $l$. Here, $k=mg/2\hbar$ is the wave number, $\hbar$ is the reduced Planck constant, the quantities $\delta_l$ represent the phase shifts of a scattered particle, while $P_l(x)$ are the Legendre polynomials of $x$. The phase shifts $\delta_l$ are determined by the speed $g$ and by the interatomic potential. The TCS defined by (\ref{CK}) is also decomposed as
\begin{equation}
\sT^{\mbox{\tiny (B)}}(g)=\f{s}{2s+1}\sigma'_{\mbox{\tiny T}}(g)+
\f{s+1}{2s+1}\sigma''_{\mbox{\tiny T}}(g),
\lae{CP}\end{equation}
\begin{equation}
\sT^{\mbox{\tiny (F)}}(g)=
\f{s+1}{2s+1}\sigma'_{\mbox{\tiny T}}(g)+\f{s}{2s+1}\sigma''_{\mbox{\tiny T}}(g),
\lae{DC}\end{equation}
for bosons and fermions in accordance with (\ref{CH}). An integration (\ref{CK}) of both $\sigma'(g,\chi)$ and $\sigma''(g,\chi)$ leads to the expressions
\begin{equation}
\sigma'_{\mbox{\tiny T}}(g)=\f{8\pi}{k^2} \sum_{l=1,3,5,...}^\infty (2l+1)\sin^2 \delta_l,
\lae{DD}\end{equation}
\begin{equation}
\sigma''_{\mbox{\tiny T}}(g)=\f{8\pi}{k^2} \sum_{l=0,2,4,...}^\infty (2l+1)\sin^2 \delta_l.
\lae{CU}\end{equation}

{
In case of distinguishable particles,  the DCS and TCS read \cite{Joa01,Lan12,Hir01,Cha04,Fer02}
\begin{equation}
\sigma(g,\chi)=\f{1}{k^2}\lt|\sum_{l=0}^\infty f_l(g,\chi) \rt|^2,
\lae{CT1}\end{equation}
and
\begin{equation}
\sigma_{\mbox{\tiny T}}(g)=\f{4\pi}{k^2} \sum_{l=0}^\infty (2l+1)\sin^2 \delta_l,
\lae{DD1}\end{equation}
respectively.}

Calculations of the phase shifts is based on  the Schr\"odinger equation \cite{Lan12,Joa01}  written down in the spherical coordinates. The method to solve this equation and to calculate the phase shifts used here is the same as that described in Ref.\cite{Sha118}.

\section{Matrix of deflection angle}

The vector $\sigma_{\mbox{\tiny T}j}$ and the matrix $\xi_{ij}$ were calculated for helium-3 and helium-4 having the atomic masses \cite{Moh03} 3.01605 u and 4.00260 u, respectively. The AI  potential for these two species is the same and can be found in Refs. \cite{Cyb01,Hel02,Prz01}. For our purpose, the potential proposed in the work \cite{Prz01} was chosen as the most complete and exact at the moment.  The authors of the paper \cite{Cen01} used this potential to calculated the viscosity and thermal conductivity of both helium-3 and helium-4. The uncertainty of these quantities caused by the potential uncertainty ranges from 0.05\% at low temperature $T$ to 0.002\% for $T>$ 50 K. Since typical numerical errors of the DSMC method are quite larger, the potential uncertainty does not contribute into a total uncertainty of numerical results obtained by the DSMC.

The non-uniform mesh composed from $N_g=800$ nodes of the speed $g$ was introduced as
\begin{equation}
g_j\mbox{(m/s)}=400\cdot(1.005^j-1).
\lae{CR}\end{equation}
The TCSs for $^3$He and $^4$He calculated for each value of $g_j$ are plotted in Figure \ref{figA} which shows their undulatory behaviours. Both isotopes $^3$He and $^4$He have practically the same TSC for large values of the speed $g$ and quite different behaviors for its small values, i.e. the TCS of $^4$He sharply increases by decreasing the speed $g$, while the TCS of $^3$He weakly varies in the same limit. At the smallest speed considered here, i.e. $g=2$ m/s, the TCS of $^4$He is four orders of magnitude larger than that of $^3$He. Such behaviours are qualitatively consistent with experimental data  \cite{Gra12}. However, a large dispersion of these data does not allow to perform a reasonable  quantitative comparison. Moreover, $N_\xi=100$ values of $\xi_{ij}$ were calculated for each speed node $g_j$ following the rule (\ref{CI}). The matrices of $\xi_{ij}$ for $^3$He and $^4$He can be requested from the author. 

The provided matrices can be used to model any flows of helium. To start calculations, the file  "xiHe3.csv" or "xiHe4.csv" is read storing the first hundred columns in the matrix $\xi_{ij}$, the 101th column is read and stored in the vector $\sigma_{\mbox{\tiny T}j}$ ($1\le j\le 800$). The quantity $(\sT g)_{\mbox{\scriptsize max}}$ is set initially to a reasonable value for each cell with a possibility to update it in case when a pair with a larger value of the product $\sT g$ arises. Let us assume a randomly chosen pair has the relative speed $g$. Then, the index $j$ is calculated as
\begin{equation}
 j= \Biggl\lfloor \f{\ln(1+g/400)}{\ln(1.005)}+\f12\Biggr\rfloor, \quad 1\le j\le 800.
\lae{CW}\end{equation}
If by chance $j>800$ (it can be happen at very high temperatures), $j$ is set equal to 800. Then the condition (\ref{CQ}) is checked. If it is true, the index $i$ is randomly chosen from the range $1\le i\le 100$. Once $i$ and $j$ are known, the element $\xi_{ij}$ is used as $\cos\chi$ to calculate the post-collision velocities according to Eqs.(8.32) - (8.35) from Ref. \cite{Sha02B}. If by chance $\sigma_{\mbox{\tiny T}j}\,g>(\sT g)_{\mbox{\scriptsize max}}$, the quantity  $(\sT g)_{\mbox{\scriptsize max}}$ is updated as $(\sT g)_{\mbox{\scriptsize max}}=\sigma_{\mbox{\tiny T}j}g$.

\begin{figure}
  \centering
  \includegraphics{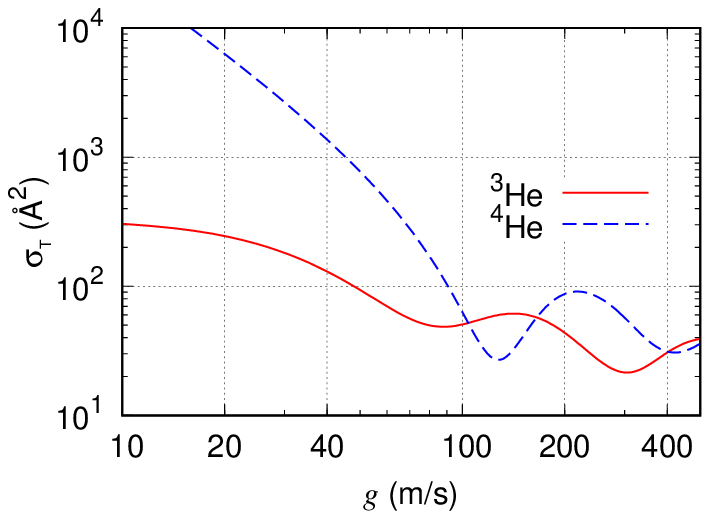}

  \includegraphics{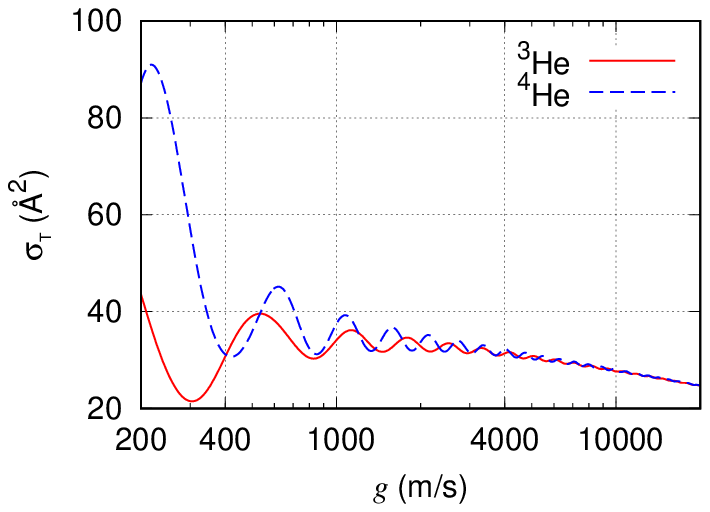}

  \caption{Total cross section $\sT$ vs relative velocity $g$}
  \lae{figA}
\end{figure}

\section{Examples}

In order to illustrate the technique and to estimate the influence of the quantum effects, two classical problems of fluid mechanics related to transport phenomena through helium were solved.

The first problem is a heat transfer between two parallel plates fixed at $x=\pm H/2$. The plate at $x=-H/2$ is kept at a temperature $T_0+\Delta T/2$, while the other plate has a lower temperature $T_0-\Delta T/2$. We are interested in the heat flux $q_x$ as a function of the gas rarefaction $\delta$ and of the equilibrium temperature $T_0$. The gas rarefaction is defined as \cite{Sha02B}
\begin{equation}
\delta=Hp_0/(\mu v_0),\quad v_0=\sqrt{2\bk T_0/m},
\lae{DH}\end{equation}
where $p_0$ is the equilibrium gas pressure, $\mu$ is its viscosity at the equilibrium temperature, $v_0$ is the most probable molecular speed. The viscosity $\mu$ reported in the papers \cite{Bic01,Cen01} for a wide range of the temperature are used to determine the rarefaction parameter. The value $\delta=1$ corresponds to the transitional regime, when the frequency of the interatomic collisions is close to that of the gas-surface collisions. The value $\delta=10$ describes the temperature jump regime, when the interatomic collisions predominate, but the gas-surface collisions are still important.

The second problem is the planar Couette flow. In other words, we consider the same plates both kept at the temperature $T_0$, but the plate at $x=-H/2$ is moving with a speed $U_w/2$ in the $y$-direction, while the plate $x=H/2$ is moving with the same speed in the opposite direction. Now, we are interested in the shear stress $P_{xy}$ between the plates in the transitional ($\delta=1$) and velocity slip regimes ($\delta=10$). In both problems, we assume the diffuse scattering of particles on the plate surfaces. The results of these two problems  will be given in terms of the dimensionless quantities defined as
\begin{equation}
Q=-q_xT_0/(p_0v_0\Delta T),\quad \Pi=-P_{xy}v_0/(p_0 U_w).
\lae{CJ}\end{equation}
The DSMC calculations were carried out dividing the space $-H/2\le x\le H/2$ into 800 cells, considering 200 particles per cell, and using the time step $\Delta t$ equal to $0.002H/v_0$. This numerical scheme provide the numerical error of $Q$ and $\Pi$  less  than 0.1\%,
estimated by carrying out test calculations with the double number of cells, the double number of particles and reducing the time step by the factor 2.

First, test calculations were carried out for $\delta=40$ in order to extract the viscosity and heat conductivity coefficients following the technique described in Ref.\cite{Sha96}. The temperature difference in the heat transfer problem was $\Delta T/T_0=0.1$ and the wall speed in the Couette flow was $U_w/v_0=0.1$. It was verified that the viscosity and thermal conductivity obtained by the DSMC method are in agreement within 0.1 \% with those reported in the works \cite{Bic01,Cen01} over the temperature range from 1 K to 3000K. Then, the heat flux problem was solved for $\Delta T/T_0=1.5$ and the Couette flow was solved for $U_w/v_0=2$. In both cases, the values $\delta=1$ and 10 were considered. The temperature $T_0$ was varied from 1 K to 3000 K. The files "xiHe3.csv" and "xiHe4.csv" were used in the quantum approach.

In order to compare this approach with that based on the classical scattering, additional two matrices were calculated following the technique described in Refs. \cite{Sha90} for the same nodes (\ref{CR}) of the speed $g$. In our calculations, the quantity $d_m$ needed for Eq.(\ref{CS}) was $3d_0$, where $d_0$ is the zero point of the potential $V(d_0)=0$, i.e. where the potential changes it own sign. The value $d_0=2.64095$ {\AA}  correspondes to the potential \cite{Prz01} used here so that the TCS is equal to $\sT=197.20$ \AA$^2$ for all values of the speed $g_j$. Note that this value is larger than the TCS based on the quantum theory for $^3$He and for $^4$He in the speed ranges $g>30$ m/s and $g>80$ m/s, respectively. In the speed range $g>400$ m/s typical in most of simulations, the quantum TCS is about 40 \AA$^2$ and even smaller. It means that the number of tested pairs calculated by Eq. (\ref{CL}) based on the quantum theory is quite smaller than that calculated by Eq. (\ref{CS}) based on the classical approach. Table \ref{tabA} contains the ratio of the computational time needed to solve the Couette problem applying the classical scattering to that using the quantum calculation. It shows that the quantum approach reduces the computational effort to simulate flows at the room temperature and higher  providing the same results as the classical approach.

\begin{table}\centering
\caption{Ratio of computational time to solve the Couette flow problem applying the classical approach $t_c$ to that using the quantum calculation $t_q$.}
\begin{tabular}{rcc}   \hline
   & \multicolumn{2}{c}{$t_c/t_q$} \\ \cline{2-3}
$T$ (K) &  $^3$He &  $^4$He \\ \hline
100     &  1.25   &    1.26  \\
300     &  1.32   &    1.37  \\
1000    &  1.49   &    1.45  \\
3000    &  1.97   &    1.59  \\
\hline
\end{tabular} \lae{tabA}\end{table}

The numerical results of the heat flux $Q$ and shear stress $\Pi$ are plotted in Figure \ref{figB} and \ref{figC}, respectively. Numerical values of $Q$ and $\Pi$ are provided in Appendix. First of all, neither difference between classical and quantum approaches nor between $^3$He and $^4$He is observed at $T_0\ge300$ K. In other words, the values of $Q$ are the same within the numerical error 0.1 \% for both isotopes and for both quantum and classical approaches. The same can be said about the shear stress $\Pi$. In the temperature range $20\le T/\mbox{K}\le 300$, the difference between the values of $Q$ and $\Pi$ based on the quantum approach and those based on the classical scattering exceed the numerical error, but still there is no difference between the gas of fermions and that of bosons. For lower temperature $T\le 20$ K, the difference between fermions $^3$He and bosons $^4$He becomes lager than the numerical error. The maximum discrepancy of the heat flux  $Q$ for the two isotopes is 6\%, while the discrepancy of the shear stress $\Pi$ reaches 18 \%. The difference of $Q$ based on the quantum scattering from those based on classical scattering reaches 60\% for $^3$He and 30 \% for $^4$He. The same differences for the shear stress $\Pi$ are 67\%  and  34\%, respectively. The qualitative behavior of the heat flux $Q$ and shear stress $\Pi$ is the same in the transitional ($\delta=1$) and hydrodynamic ($\delta=10$) regimes.

\begin{figure}
  \centering
  \includegraphics{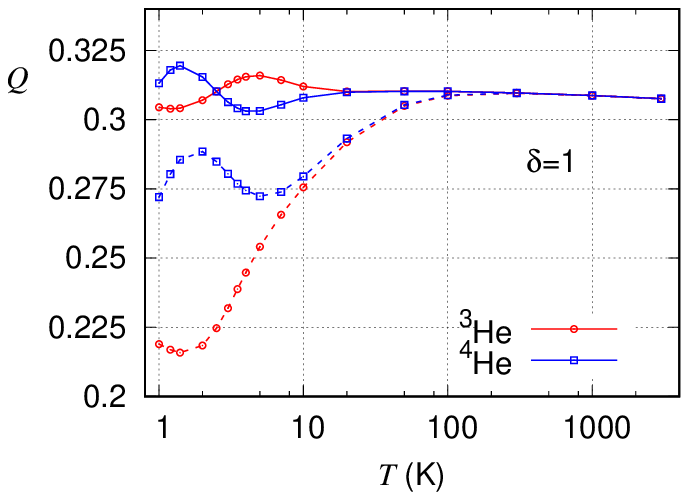}
  \includegraphics{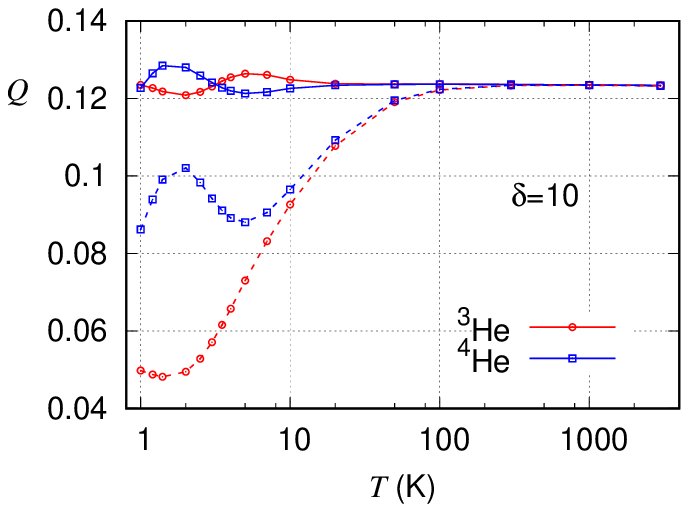}
  \caption{Reduced heat flux $Q$ vs temperature $T$: solid lines - quantum scattering, dashed lines - classical scattering}
  \lae{figB}
\end{figure}

\begin{figure}
  \centering
  \includegraphics{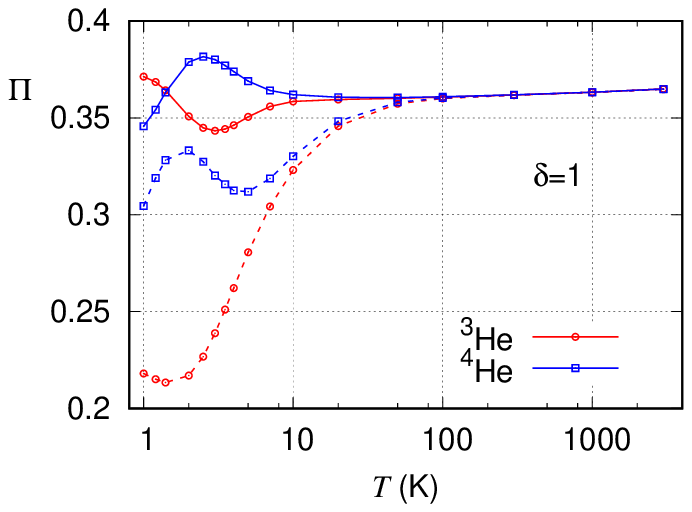}
  \includegraphics{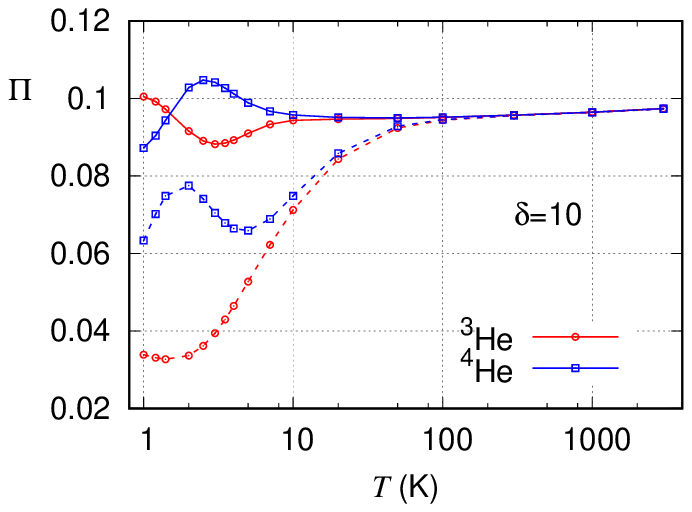}
  \caption{Reduced shear stress $\Pi$ vs temperature $T$: solid lines - quantum scattering, dashed lines - classical scattering}
  \lae{figC}
\end{figure}

{
Usually, measurements of the heat transfer \cite{Sem01,Tro01} and torque in Couette flow \cite{Sta01,Alo01} are done at a temperature close to the ambient one using the isotope $^4$He. Then, the same behaviours of these phenomena are assumed for any temperature and for any isotope. The above reported results show that the behavior of the transport phenomena vary from one isotope to another at low temperatures. Therefor, we encourage experimentalists to perform measurements of such phenomena at low temperature with both isotopes $^3$He and $^4$He.}

\section{Conclusions}

An interatomic interaction based on quantum scattering was implemented into the direct simulation Monte Carlo method applied to transport phenomena through rarefied gases. Such an implementation allows us to model  flows of light gases like helium over the whole temperature range beginning from 1 K up any temperature when no ionization happens. As an example, two helium isotopes $^3$He and $^4$He have  been considered  in two classical problems of fluid mechanics, namely, heat transfer between two planar surfaces and planar Couette flow. The problems have been solved over the temperature range from 1 K to 3000 K and for two values of the rarefaction parameter $\delta=1$ and 10. The former corresponds to  the transitional regime and the latter describes the temperature jump and velocity slip regime. The heat flux and shear stress were calculated with the numerical error less than 0.1\%. No influence of the quantum effects was detected within this error for the temperature 300 K and higher. However, the quantum approach requires less computational effort than the classical one in this range of the temperature because the quantum total cross section is relatively small, while it is not well determined in the frame of classical theory.  For temperatures lower than 300 K, the influence of the quantum effects exceed the numerical error and increases by decreasing the temperature. The behaviours of fermions $^3$He and bosons $^4$He are qualitatively different at a temperature lower than 20 K. The difference between the quantum and classical approaches can reach 67 \% for the problems considered here. The matrices of the deflection angle for $^3$He and $^4$He calculated for the present paper can be used to model any flow of these gases. It should be noted that the influence of quantum effect can be larger for flows with a larger temperature variation, e.g., supersonic flows. Usually, flows of gaseous mixtures \cite{Sha100,Sha105} are more sensitive to intermolecular potential than single gas flows so that the influence of quantum effect in case of mixtures also can be larger.

\section*{Acknowledgments}

The author acknowledges the Brazilian Agency CNPq for the support of his research, grant 303697/2014-8.

\appendix

\section{Numerical data for Figures \ref{figB} and \ref{figC}}

\begin{table}[h]
\caption{Dimensionless heat flux $Q$ vs temperature and rarefaction parameter $\delta$}
\begin{tabular}{rcccccccc}  \hline
& \multicolumn{8}{c}{$Q$}\\ \cline{2-9}
&\multicolumn{4}{c}{$\delta=1$} &\multicolumn{4}{c}{$\delta=10$}\\
&  \multicolumn{2}{c}{quantum}&  \multicolumn{2}{c}{classical}&
 \multicolumn{2}{c}{quantum}&  \multicolumn{2}{c}{classical}\\
$T$(K) & $^3$He & $^4$He & $^3$He & $^4$He & $^3$He & $^4$He & $^3$He & $^4$He \\ \hline
1~~	  &0.30447	&0.31322	&0.21890  &0.27203 & 0.12351	&0.12274	&0.04976	&0.08622  \\
1.2	  &0.30398	&0.31800	&0.21691  &0.28030 & 0.12268	&0.12649	&0.04873	&0.09391    \\
1.4	  &0.30419	&0.31954	&0.21579  &0.28552 & 0.12182	&0.12846	&0.04817	&0.09908    \\
2~~	  &0.30707	&0.31547	&0.21838  &0.28846 & 0.12085	&0.12805	&0.04944	&0.10206    \\
2.5	  &0.31017	&0.31027	&0.22468  &0.28484 & 0.12171	&0.12591	&0.05282	&0.09831        \\
3~~	  &0.31282	&0.30637	&0.23189  &0.28044 & 0.12312	&0.12413	&0.05708	&0.09418        \\
3.5	  &0.31454	&0.30414	&0.23876  &0.27687 & 0.12447	&0.12278	&0.06154	&0.09109        \\
4~~	  &0.31553	&0.30313	&0.24477  &0.27440 & 0.12549	&0.12198	&0.06572	&0.08919        \\
5~~	  &0.31599	&0.30318	&0.25411  &0.27233 & 0.12642	&0.12133	&0.07299	&0.08805        \\
7~~	  &0.31435	&0.30540	&0.26563  &0.27388 & 0.12609	&0.12168	&0.08315	&0.09056        \\
10~~  &0.31200	&0.30791	&0.27557  &0.27952 & 0.12486	&0.12259	&0.09260	&0.09651        \\
20~~  &0.31018	&0.30992	&0.29184  &0.29317 & 0.12380	&0.12340	&0.10776	&0.10927        \\
50~~  &0.31034	&0.31028	&0.30499  &0.30535 & 0.12372	&0.12365	&0.11908	&0.11951        \\
100~~ &0.31027	&0.31023	&0.30871  &0.30885 & 0.12369    &0.12368	&0.12222	&0.12237        \\
300~~ &0.30970  &0.30969	&0.30956  &0.30961 & 0.12360	&0.12362	&0.12339	&0.12343         \\
1000~~&0.30878	&0.30874	&0.30872  &0.30874 & 0.12347	&0.12349	&0.12343	&0.12344        \\
3000~~&0.30762	&0.30763	&0.30753  &0.30756 & 0.12337	&0.12337	&0.12329	&0.12328        \\   \hline
\end{tabular}\lae{tabAA} \end{table}

\begin{table}
\caption{Dimensionless shear stress $\Pi$ vs temperature and rarefaction parameter  $\delta$}
\begin{tabular}{rcccccccc}  \hline
& \multicolumn{8}{c}{$\Pi$}\\ \cline{2-9}
&\multicolumn{4}{c}{$\delta=1$} &\multicolumn{4}{c}{$\delta=10$}\\
&  \multicolumn{2}{c}{quantum}&  \multicolumn{2}{c}{classical}&
 \multicolumn{2}{c}{quantum}&  \multicolumn{2}{c}{classical}\\
$T$(K) & $^3$He & $^4$He & $^3$He & $^4$He & $^3$He & $^4$He & $^3$He & $^4$He \\ \hline
1~~	&0.37127&	0.34564&	0.21805&	0.30454  &0.10049&	0.08723&	0.03386&	0.06333\\
1.2 &0.36855&	0.35421&	0.21509&	0.31899  &0.09919&	0.09040&	0.03310&	0.07014 \\
1.4 &0.36419&	0.36314&	0.21334&	0.32823  &0.09725&	0.09438&	0.03268&	0.07485 \\
2~~	&0.35080&	0.37883&	0.21698&	0.33325  &0.09158&	0.10285&	0.03361&	0.07753\\
2.5 &0.34490&	0.38162&	0.22674&	0.32738  &0.08905&	0.10476&	0.03612&	0.07410 \\
3~~	&0.34330&	0.38008&	0.23885&	0.32019  &0.08821&	0.10417&	0.03940&	0.07047\\
3.5 &0.34425&	0.37702&	0.25100&	0.31570  &0.08847&	0.10271&	0.04294&	0.06789 \\
4~~	&0.34619&	0.37394&	0.26215&	0.31254  &0.08923&	0.10121&	0.04643&	0.06643\\
5~~	&0.35053&	0.36896&	0.28056&	0.31189  &0.09100&	0.09888&	0.05274&	0.06588\\
7~~	&0.35592&	0.36414&	0.30423&	0.31867  &0.09331&	0.09671&	0.06221&	0.06892\\
10~~  &0.35851&	0.36195&	0.32306&	0.33014  &0.09436&	0.09576&	0.07122&	0.07485 \\
20~~  &0.35952&	0.36065&	0.34575&	0.34824  &0.09466&	0.09512&	0.08435&	0.08585 \\
50~~  &0.36012&	0.36051&	0.35734&	0.35803  &0.09486&	0.09500&	0.09238&	0.09282 \\
100~~ &0.36072&	0.36088&	0.36002&	0.36029  &0.09513&	0.09518&	0.09441&	0.09456 \\
300~~ &0.36184&	0.36188&	0.36171&	0.36183  &0.09570&	0.09571&	0.09561&	0.09561 \\
1000~~&0.36321&	0.36329&	0.36318&	0.36321  &0.09649&	0.09647&	0.09642&	0.09644 \\
3000~~&0.36491&	0.36495&	0.36474&	0.36479  &0.09742&	0.09742&	0.09731&	0.09732 \\ \hline
\end{tabular} \lae{tabBB}\end{table}


\end{document}